\documentclass[11pt,a4paper]{article}

\usepackage[
top    = 2.50cm,
bottom = 2.50cm,
left   = 2.50cm,
right  = 2.50cm]{geometry}
\usepackage{jheppub}
\usepackage[sort&compress]{natbib}

\usepackage{amssymb}
\usepackage{graphicx}
\usepackage{slashed}
\usepackage{amsmath}
\usepackage{hyperref}
\usepackage{verbatim}

\usepackage{caption}
\usepackage{subcaption}

\title{Higher Dimensional Generalizations of the SYK Model}

\author[a]{Micha Berkooz}
\author[b]{\!, Prithvi Narayan}
\author[c]{\!, Moshe Rozali}
\author[d]{\!, Joan Sim\'on}
\affiliation[\,a]{
\it{Department of Particle Physics and Astrophysics,\\
Weizmann Institute of Science, Rehovot 7610001, Israel}}
\affiliation[\,b]{
\it{International Centre for Theoretical Sciences, Hesaraghatta,\\ 
Bengaluru North, 560 089, India}}

\affiliation[\,c]{
\it{
Department of Physics and Astronomy, University of British Columbia,\\
Vancouver, BC V6T 1Z1, Canada}}

\affiliation[\,d]{
\it{School of Mathematics and Maxwell Institute for Mathematical Sciences,\\
        University of Edinburgh, King's Buildings,
        Edinburgh EH9 3FD, UK}}
\emailAdd{Micha.Berkooz@weizmann.ac.il}
\emailAdd{prithvi.narayan@gmail.com}
\emailAdd{rozali@phas.ubc.ca}
\emailAdd{j.simon@ed.ac.uk}

\abstract{We discuss a 1+1 dimensional generalization of the Sachdev-Ye-Kitaev model. The model contains $N$ Majorana fermions at each lattice site with a nearest-neighbour hopping term. The SYK random interaction is restricted to low momentum fermions of definite chirality within each lattice site. This gives rise to an ordinary 1+1 field theory above some energy scale and a low energy SYK-like behavior. We exhibit a class of low-pass filters which give rise to a rich variety of hyperscaling behaviour in the IR. We also discuss another set of generalizations which describes probing an SYK system with an external fermion, together with the new scaling behavior they exhibit in the IR. 
}

\begin{document}

\maketitle

\section{Introduction and Conclusions}
\label{sec:Introduction}

The Sachdev-Ye-Kitaev (SYK) model \cite{Sachdev:1992fk,KitaevTalks} has received much attention recently \cite{Polchinski:2016xgd,Maldacena:2016hyu,Jevicki:2016bwu,Fu:2016yrv,You:2016ldz,Jevicki:2016ito}. It is a simple model with solvable aspects which exhibits interesting connections to quantum chaos \cite{Shenker:2013pqa, Shenker:2013yza,Shenker:2014cwa,Maldacena:2015waa,Reynolds:2016pmi}, black hole physics and quantum gravity in 1+1 dimensions 
\cite{Almheiri:2014cka,Sachdev:2015efa,Maldacena:2016upp,Jensen:2016pah,Sekino:2008he,Engelsoy:2016xyb,Cvetic:2016eiv}. Understanding these relations better in this model, and potential extensions of it, is an active and exciting research direction that promises to improve our knowledge of holography. 

The SYK model is a quantum mechanical model involving Majorana fermions interacting with non-local random couplings. Much of the interesting physics of the model, including low energy near-conformal symmetry\footnote{The relation to $AdS_2$ was first pointed out in \cite{PhysRevLett.105.151602}.}and maximal scrambling, is not manifestly related to the specific construction in a transparent manner. To gain a better understanding of these features and their origin, it is therefore important to explore generalizations of the model. There has already been some work in this direction\footnote{In \cite{Gu:2016oyy}, a 1+1 translationally invariant model (on average), and higher dimensional extensions, were proposed. In \cite{Gross:2016kjj}, the authors propose a generalization by introducing an extra flavour index for fermions and considering complicated interactions between them. These models have neither hopping term nor a low momentum filter. Hence, we believe our models are qualitatively different from theirs. Another generalization including hopping term (but not low momentum filter) was introduced in \cite{PhysRevB.59.5341}. However the details of the interaction are different from ours.}.

Our approach to exploring the generalization to the SYK model is motivated by holography. In this work we consider two models. In the first one, we consider a 1+1d generalization in which we add locality in the extra dimension and implement random couplings as a function of the momentum. In the second model, we consider probe fermions interacting with a core of SYK degrees of freedom.

The first extension consists of a chain of SYK models, with Majorana fermions at each site with nearest  hopping. This adds local physics in the added spatial direction. The fermions interact via random couplings, as in the original SYK model, but only at low enough momenta. This is achieved by first passing the Majorana fermions via a low-pass filter, and then coupling these via an SYK random interaction. The motivation for this is to have a model which is an ordinary relativistic field theory above some scale (and below a UV cut-off), which may model an asymptotically $AdS_3$ space, and some complicated IR dynamics encoding an object in the interior of $AdS_3$. Besides ensuring that the high momenta modes are filtered out, in this work we also focus on {\it chiral} filters, thus only one chiral half of the fermions participate in the interactions.

The second extension considers a core of SYK fermions to which a probe fermion is coupled to. In this approach we interpret the SYK fermions as describing the interior of the black hole, and the probe fermions as a single trace operator outside of it. 

The results we obtain for the first class of models demonstrate, depending on the precise way the low pass filter is implemented, a rich variety of IR theories generalizing the SYK family of models. At high momentum, the model asymptotes to a free 1+1 dimensional field theory. As we decrease the momentum, the modes interact more strongly, until the new scaling regime is approached. In this scaling regime the fermions acquire an anomalous dimension, with different scaling for the space and time coordinate.  In other words, we get a general hyperscaling at low energies. The dynamical critical exponent $z$ depends on the type of low-pass filter we use, and we discuss the range of sensible possibilities that arise. We also solve an example of the second class of models and discuss the new scaling dimensions appearing for these fermions.

One could consider our models as a particular class of disordered large $N$ theories at strong coupling. Applications of holography to such theories have been explored in \cite{Aharony:2015aea}. Furthermore, that inherent randomness in the disordered theory might be crucial for understanding black hole physics has recently been pointed out \cite{Balasubramanian:2014gla}. Interesting connections between 1+1d theories and black holes in gravity were also previously explored \cite{Guica:2008mu,Berkooz:2006wc,Berkooz:2014uwa}. 

The outline of our paper is as follows. In section 2 we introduce the model of interacting fermions on a discrete lattice and the associated low-pass filters. We solve for the two point function in section 3 for scaling filters, exhibit different deep IR scaling dimensions, and discuss the continuum limit. In section 4 we discuss the probe fermion models and solve one such example.  In appendix A, we derive the Schwinger-Dyson equations using the replica formulation and in appendix B we discuss gaussian filters.

While we focus below on simple 1+1 dimensional chains, it is possible to extend the model to several spatial directions and various interesting lattice structures in those directions.  More generally, we view this work as a preliminary study of a large set of models generalizing the SYK construction both in the UV and the IR. It would be interesting to further study those models, compare and contrast their features with those of the original SYK model. In particular, studying the 4-point function would allow us to probe the chaotic behaviour of the system and the spatial spread of chaos, as manifested by the butterfly velocity. It might also be interesting to compute entanglement entropy (perhaps numerically as in \cite{Fu:2016yrv}). More ambitiously, one can hope that subtle issues like the information paradox \cite{Mathur:2010kx,Almheiri:2012rt} might be clearer if one has solvable models capturing the relevant physics of higher dimensional versions of the AdS/CFT correspondence.




\section{The 1+1D "low pass" SYK model}

\subsection{Definition of the Model}

Consider an extension of the SYK model involving a one-dimensional lattice with $L$ sites having $N$ Majorana fermions $\chi^{i,a}$ on each site $i$ with an
 $\mathrm{SO}(N)$ index $a$. Its euclidean space lagrangian is
\begin{equation}
\mathcal{L}_E=\sum_{i,a}\left\{ \frac{1}{2}\chi^{i,a}\partial_{\tau}\chi^{i,a} - i\alpha[\chi^{i,a},\chi^{i+1,a}] \right\} +\sum_{i,abcd}J_{i,abcd}\eta^{i,a}\eta^{i,b}\eta^{i,c}\eta^{i,d}\,.
\label{trans-inv}
\end{equation}
We can either take the lattice to be a discretized circle or a discretized infinite line. The free theory involves a hopping term with bare parameter $\alpha$. The interaction term involves random couplings satisfying the disorder average
\begin{equation}
\langle J_{iabcd}J_{jabcd}\rangle=\frac{3!J^{2}}{N^3} \delta_{ij}  \hspace{10mm} \mbox{no sum over }a,b,c,d
\label{dis-av}
\end{equation}
and \textit{low pass} fermions $\eta^{i,a} $ defined by a filter function $F$
\begin{equation}
\eta^{i,a}=\sum_{j} {\tilde F}(i-j)\chi^{j,a}\,.
\label{lowp}
\end{equation}
We will be interested in two filters : the standard {\it gaussian} filter
\begin{equation}
{\tilde F}(i-j)={\tilde A} \, \exp({-\hat{D}^{2}\frac{(i-j)^{2}}{L^{2}}})\,,
\end{equation}
where $\hat D$ sets the scale of the filter, and the {\it "scaling"} filters
\begin{equation}
{\tilde F}(i-j) \sim {{\tilde A}\over |i-j|^{\gamma\prime}}\label{filtpos}
\end{equation}
for a large enough range of $|i-j|$ and an appropriate range of $\gamma^\prime$. Depending on the latter, we may need to soften the filter at short distances or provide a sharper cut-off at large distances. We will assume the filter behavior is as in \eqref{filtpos} for a large enough range of lattice site separations, and discuss potential UV and IR modifications when we need it.

As mentioned in the introduction, some of the features of this model attempt to resemble the physics of $AdS_3$ black holes :
\begin{itemize}
        \item At high momenta, the $\chi$ fermions decouple from the random interaction and  become free -- this is the analogue of the region of $AdS_3$ far from the black hole. In the intermediate regime, where momentum is larger than the scale of the low pass filter but still smaller then the inverse lattice spacing, the model describes the simplest conformal field theory -- N species of Majorana fermions -- and, at least kinematically, we can think about it as $AdS_3$. We could of course also complicate the theory further in that regime, but in this work we keep the UV theory free (in section 5 we will discuss another interpretation where the free modes are modes outside the horizon of single trace operators).
        \item  At low momenta, the $\chi$ quanta become strongly interacting, in the appropriate SYK limit $J\rightarrow\infty$ -- this is akin to low momentum modes of the dual field theory forming a plasma, encoded by a dual black hole. 
\end{itemize}

The model is not quite SYK -- other than the zero modes, all other modes are gapped in a specific pattern (if we think about them as quantum mechanics). The number of interacting fermions first increases, as a function of their momentum, as more and more modes participate in the interaction. It then decreases when the cut-off of the filter is reached. This will bring about different IR scaling behaviors, depending on the shape of the smearing function. We will see below that we obtain a large set of models with distinct scaling behaviors for the time coordinate and for the spatial coordinate.

The specific model that we will discuss is a chiral theory. As is standard with lattice fermions, the free model flows in the IR to a non-chiral theory of Majorana fermions. However, the low pass filter that we defined above keeps the low momentum modes of only one chirality of the fermions. The low-pass fermion for the other chirality is defined as (for the gaussian filter as an example)
\begin{equation}
\eta^{R,i,a}={\hat A}\sum_{j}e^{-\hat{D}^{2}\frac{(i-j)^{2}}{L^{2}}}(-1)^{i-j}\chi^{j,a}
\end{equation}
The present model includes only interactions of the low $\eta^L$ and not of low momenta $\eta^R$. We will refer to it as the chiral model -- similar models where interactions involve both left and right can be made non-chiral but here we will discuss only the former.

\subsubsection{Possible generalizations}

Several interesting generalizations are possible for the model discussed above.  
\begin{itemize}
        \item  One possible generalization is the non-chiral model just mentioned. In this generalization, there will be a random interaction for the right movers $\eta^{R,i,a}$ as well. This theory will preserve parity symmetry (discussed in some detail in section \ref{sec:FreeTheory}). We can also add a direct coupling between the left and right moving sectors of the form $J_{i,abcd}\eta^{L,i,a}\eta^{L,i,b}\eta^{R,i,c}\eta^{R,i,d}$.
        
     \item  The random interactions in \eqref{trans-inv} involve 4 Majorana fermions. More generally, as was done in the  0+1 dimensional SYK model \cite{Maldacena:2016upp}, one can have a random interaction involving an arbitrary number $q$ of fermions. In that case, the theory is exactly solvable for $q=2$ and has important implications for large values of $q$ that allow to obtain an analytic understanding of the entire flow. We expect the same to hold here.
     
      \item
      The model \eqref{trans-inv} is not translationally invariant because the interactions $J_{i,abcd}$ depend on the lattice site $i$, although correlation functions are translationally invariant after the disorder average. We can modify the model above to accommodate strict translational invariance replacing the interaction term by, for example,  
    \begin{equation}
        \mathcal{L}_{\text{int}} = \sum_i \sum_{a,b,c,d}\sum_{d_1,d_2,d_3} J_{abcdd_1d_2d_3} \eta^{i,a}\eta^{i+d_1,b}\eta^{i+d_2,c}\eta^{i+d_3,d}
        \end{equation}
        This model is not easily solvable using the tools we discuss below, but we hope to return to it in the near future.
       
\end{itemize}

\subsection{The free theory}\label{sec:FreeTheory}

Using euclidean conventions (with $Z=e^{-S_E},\ S_E=\int d\tau {\mathcal L}_E, \ d\tau=idt$), the lagrangian is
\begin{equation}\label{EuclideanFree}
{\mathcal L}^{\text{free}}_E=\sum_{i,a} \left\lbrace {1 \over { 2}} \chi^{i,a}\partial_\tau \chi^{i,a} -  i\alpha [\chi^{i,a},\chi^{i+1,a}] \right\rbrace \, .
\end{equation}
To describe a periodic lattice, we identify $\chi^{i,a}=\chi^{i+L,a}$. Hence, as operators, the commutation relations are
\begin{equation}
\{\chi^{i,a},\chi^{j,b}\}=\delta^{ab}\sum_{p\in \mathbb{Z}}\delta^{i(j+pL)}\,.
\end{equation}
The momentum space fermions $\chi_k^a$ are defined, for integer $k$, as 
\begin{equation}
\chi_k^a = \frac{1}{\sqrt{L}}\sum_j e^{2\pi i {jk\over L}}\, \chi^{j,a}\, \quad \text{with} \quad \chi^{j,a}= \frac{1}{\sqrt{L}} \sum_k e^{-2\pi i {jk\over L}}\,\chi_k^a\,. 
\label{mfermion}
\end{equation}
The conventions will be that momentum index is down and position index is up. 

We assume that L is even. Since $\chi_k^a=\chi_{k+L}^a$, we can either take $k$
to be an arbitrary integer with this periodicity, or we can restrict ourselves to the range
\begin{equation}
k= -L/2+1,...L/2
\end{equation}
We will use both these descriptions.  As operators, the momentum space fermions satisfy the commutation relations
\begin{equation}\label{anticom}
\{\chi_k^a,\chi_{k'}^b\} = \delta^{ab}(\delta_{k+k',0}+\delta_{k+k',L} )\,,
\end{equation}
where the second contribution is only non-zero when $k=k'=L/2$. The free theory \eqref{EuclideanFree} in momentum space modes becomes\footnote{We have dropped some non-essential constant pieces in obtaining this expression.} 
\begin{equation}\label{EuclideanFreeMomentum}
 {\mathcal L}^{\text{free}}_E = \sum_a \left\lbrace  \sum_{k=1}^{{L \over 2}-1}  \chi^a_{-k} ( \partial_\tau + E_k) \chi^a_k + { \chi^a_0 \partial_\tau \chi^a_0 + \chi^a_{L \over 2} \partial_\tau \chi^a_{L \over 2} \over 2}    \right\rbrace
\end{equation}
where 
\begin{equation}\label{Energy}
E_k \equiv  4 \alpha \left| \sin\left({2\pi k \over L}\right) \right|\,.
\end{equation}
Since $(\chi^a_k)^\dagger = \chi_{-k}^a$, the fermions $\chi^{a}_k$ for $1 \le k < {L \over 2}$ can be thought of as complex fermions, whereas $\chi^a_0 = {\chi^a_0}^\dagger,\chi^a_{L \over 2}={\chi^a_{L \over 2}}^\dagger$ are $2N$ Majorana fermions.

\subsubsection*{Linearized theory} 
The free theory \eqref{EuclideanFree} is non-chiral since it is invariant under the parity transformation
\begin{equation}
\chi^{i,a}\to (-1)^i \chi^{L-i,a}  
\end{equation}
Its action in momentum space is
\begin{eqnarray}\nonumber
 \chi^a_k \leftrightarrow  \chi^a_{ {L \over 2} - k} \,, \,\,\chi^a_{-k} & \leftrightarrow &   \chi^a_{- ({L \over 2} - k)}\hspace{10mm} \mbox{ for } 1 \le k < {L \over 2} \\
\chi^a_0 &\leftrightarrow & \chi^a_{L \over 2}  
\end{eqnarray}

If we define\footnote{We will take $L$ not divisible by $4$ for simplicity. $[x]$ below denotes the integer part of $x$.}
\begin{equation}\label{LRfermions}
\chi^{L,a}_k \equiv \chi_k^a,\ \ \ \chi^{R,a}_{-k} \equiv \chi_{{L \over 2}-k}^a \mbox{ for }   1 \le k \le  [{L \over 4}] \,,
\end{equation}
parity maps left $\chi^{L,a}_k$ to right $\chi^{R,a}_{-k}$ fermions. Using these degrees of freedom, the action \eqref{EuclideanFreeMomentum} becomes 
\begin{equation}\label{FreeMomentumDiscrete}
{\mathcal L}^{\text{free}}_E = \sum_a \left\lbrace  \sum_{k=1}^{ [{L \over 4}]}  \left[ \chi^{L,a}_{-k} ( \partial_\tau + E_k) \chi^{L,a}_k + \chi^{R,a}_{k} ( \partial_\tau + E_k) \chi^{R,a}_{-k} \right] + { \chi^a_0 \partial_\tau \chi^a_0 + \chi^a_{L \over 2} \partial_\tau \chi^a_{L \over 2} \over 2}    \right\rbrace\,.
\end{equation}
This form of the action will is more useful when taking the continuum limit $L\to \infty$ and linearising the dispersion relation.

\subsubsection*{States of the free theory}

The ground states of the free theory $ |\beta \rangle $ satisfy
\begin{equation}
\chi_k^a | \beta\rangle  = 0 \mbox{ for } 1 \le k < L/2
\end{equation}
since $\chi_{k}^a$ are annihilation operators for $k>0$ and creation operators for $k<0$. Hence, they form $2^{N}$ dimensional representations of $\chi^a_0, \chi^a_{L \over 2}$. These states can equivalently be described in the left and right representation as
\begin{equation}
\chi_k^{L,a}|\beta\rangle=\chi_{-k}^{R,a}|\beta\rangle=0,\ \ \ k>0
\end{equation}
Excited states are of the form
\begin{equation}
\chi^a_{-k} | \beta \rangle \mbox{ for } 1 \le k < L/2 
\end{equation}  
They carry energy $E_k$ as in \eqref{Energy}.

\subsubsection*{Momentum  space correlators}

We now compute the time ordered propagator for the momentum space fermions in any of the vacuum states
\begin{equation}
G^{ab}_{k,k'}(\tau) \equiv \langle  T\left[ \chi^a_k(\tau) \chi^b_{-k}(0)  \right] \rangle\,.
\end{equation}
Using standard methods, we obtain
\begin{eqnarray}
\langle T\left[ \chi^a_k(\tau) \chi^b_{-k}(0)  \right]   \rangle &=&  \theta(\tau) e^{-E_k \tau} \delta_{ab} \mbox{ for } 1 \le k <{ L \over 2}\,,         \\
\langle T\left[ \chi^a_{-k}(\tau) \chi^b_{k}(0)  \right]   \rangle &=&   -\theta(-\tau) e^{-E_k |\tau|} \delta_{ab} \mbox{ for } 1 \le k <{ L \over 2}\,,   \\
\langle T\left[ \chi^a_0(\tau) \chi^b_{0}(0)  \right]   \rangle &=&  \langle T\left[ \chi^a_{L \over 2}(\tau) \chi^b_{L \over 2}(0)  \right]  = \frac{1}{2}\mbox{sgn}(\tau)\,. e^{-\epsilon |\tau|} \delta_{ab} 
\end{eqnarray}
Here $\epsilon$ is small and positive and constitutes some effective "$\epsilon$ prescription".Correlators of left and right fermions in equal
\begin{equation}
\langle T\left[ \chi^{L,a}_k(\tau) \chi^{L,b}_{-k}(0)  \right]   \rangle =  \langle T\left[ \chi^{R,a}_{-k}(\tau) \chi^{R,b}_{k}(0)  \right]   \rangle=   \theta(\tau) e^{-E_k \tau} \delta_{ab}\,.
\end{equation}

It is possible to assemble all these propagators in a more compact notation 
\begin{equation}
G^{ab}_{k,k'}(\tau)  = \langle T\left[ \chi^a_k(\tau) \chi^b_{k'}(0)  \right]   \rangle = \delta_{k+k'=0,L} \delta_{ab}
\left[
\theta(\tau) e^{-E_k \tau} H(k) - \theta(-\tau) e^{-E_{k'} |\tau|} H(k')    \right] 
\end{equation}
introducing the function
\begin{equation}
H(k) = \begin{cases} 
1 &, \mbox{ if } 1 \le k <{ L \over 2}\\
0 &, \mbox{ if } -{L \over 2} < k \le 1 \\
{1 \over 2} &, \mbox{ if } k =0,{L \over 2}  \\
\end{cases}
\end{equation}
Here $E_0 = E_{L/2} =  \epsilon$ which is a regularization prescription. We have also used
$E_k = E_{-k} = 4 \alpha |\sin({2 \pi k \over L})|$. 

It is also possible to write the correlators in frequency space $(\omega)$
\begin{equation}
G^{ab}_{k,k'}(\omega) \equiv \int_{-\infty}^\infty d\tau e^{i\omega \tau} G^{ab}_{k,k'}(\tau)  = \delta_{k+k'=0,L} \delta_{ab}
 \left[
{H(k) \over -i \omega + E_k } + { H(k') \over -i \omega - E_{k'}  }     \right] \,.
\end{equation}

\subsubsection*{Continuum limit of the free theory}

We describe the continuum of the free theory here, since it would be relevant when we solve the interacting theory below. We consider the theory on a circle of size $R$. Hence, we define the continuum limit as $L \to \infty$ keeping the coordinate $x = {i \over L} R$ fixed. The physical momentum modes are $p = {2 \pi k \over R}$ with $k =0,\pm 1,\dots$ and the UV cutoff is $\Lambda \ll  {L \over R}$. 

Linearizing the energy spectrum in \eqref{Energy} for small $k$, one gets $E_p = E_k  \approx {4 \alpha R \over L} p$. We will choose the bare parameter $\alpha = {L \over 4R}\equiv {\Lambda_0 \over 4}$ from now onwards to obtain a relativistic theory, but we could have accommodated other values. Since we will be interested in chiral models, we give here the chiral sector of the free theory \eqref{EuclideanFreeMomentum} in the above limit
\begin{equation} 
{\mathcal L}^{\text{free}}_E = \sum_a \left\lbrace  \sum_{p={2 \pi \over R}}^{\Lambda}    \chi^a_{-p} ( \partial_\tau + p)  \chi^a_p  +  {1 \over 2}   \chi^a_0 \partial_\tau  \chi^a_0     \right\rbrace\,.
\label{freeR}
\end{equation}
\subsection{The interacting theory}

The interacting theory in momentum space equals 
\begin{equation}
\begin{aligned}
\mathcal{L}_E &=\mathcal{L}^{\text{free}}_E + \mathcal{L}^{\text{int}}_E = \sum_{k}\chi_{-k}^{a}(\partial_{\tau }+E_{k})\chi_{k}^{a} \\
& + \frac{1}{L^2}\sum_{a,b,c,d,k_{1},k_{2},k_{3},k_{4}}\eta_{k_{1}}^{a}\eta_{k_{2}}^{b}\eta_{k_{3}}^{c}\eta_{k_{4}}^{d}
\left(\sum_{j}e^{-2\pi i{(k_{1}+k_{2}+k_{3}+k_{4})j\over L}}J_{j,abcd}\right)
\end{aligned}
\label{model}
\end{equation}
where the low pass momentum fermions $\eta^a_k$ are defined as in \eqref{mfermion}. These are related to the physical fermions $\chi_k^a$ through the low pass filter in momentum space $F(k)$ by
\begin{equation}
\eta^i_k = A\,F(k) \chi^i_k\,.
\end{equation}
where $A$ is a normalization constant and $F(k)$ is the Fourier transform of $\tilde F$ introduced before. This allows to write the interaction lagrangian in terms of the physical fermions $\chi_k^a$
\begin{equation}
  {\mathcal L}^{\text{int}}_E =  {A^4\over L^2}\sum_{k_{1,2,3,4}} \sum_{a,b,c,d} \left(\prod_{h=1}^4 F(k_h)\right) \chi_{k_1}^a \chi_{k_2}^b \chi_{k_3}^c \chi_{k_4}^d 
\biggl(\sum_j J_{j,abcd}\,e^{-i2\pi {j\sum_{l=1}^4 k_l \over L}}\biggr)
\label{Lintchimomenta}
\end{equation}
Remember that our convention for the impurity average will be
\begin{equation}
E[J_{iabcd} J_{jabcd}] = {3!J^2 \over N^3} \delta_{ij}\,,
\label{impurity}
\end{equation}
and we will mainly be interested in scaling filters of the form
\begin{equation}
F(k)= |k|^{-\gamma}\,,
\end{equation}
or in gaussian filters
\begin{equation}
F(k)=  e^{-\pi^2k^2\over {\hat D}^2}\,.
\end{equation}

\section{Solution of the model}

In this section we find a saddle point solution to our model \eqref{model} with a scaling filter, generalizing the SYK model one. Gaussian filters are discussed in appendix \ref{sgaussian}. We use the saddle point equations to calculate the two-point functions 
\begin{equation}
  G^{ab}_{k',k}(\omega)=\langle [\chi^a_{k'} \chi^b_{k}]\rangle(\omega)\,,
\end{equation}
in an scaling regime at low energies.  We will find a rich variety of behaviour for these correlators. 

\subsection{Recalling SYK}

We start by briefly recalling the results in the original 0+1 dimensional SYK model
\begin{equation}
H_{SYK}=   \sum_{abcd} J_{abcd} \chi^a\chi^b\chi^c\chi^d\,.
\end{equation} 
where $J_{abcd}$ is the random coupling. The two-point function for the free theory is  
\begin{equation}
G_{\text{free}}(\tau)={1\over 2}\mbox{sgn}(\tau),\ \ \ \ G_{\text{free}}(w)=\int dt e^{iwt} G_{\text{free}}(\tau) = -{1\over iw}\, .
\end{equation}
In the interacting theory the connected 2 point functions are \cite{KitaevTalks,Polchinski:2016xgd,Maldacena:2016hyu} 
\begin{equation}
G_c(\tau) \sim{ 1\over  |\tau|^{2\Delta} } \mbox{sgn}(\tau)~~~~~~~\Delta=\frac{1}{4}
\end{equation}
where to transform to frequency space the following Fourier transform formula can be used
\begin{equation}
\int_{-\infty}^\infty d\tau e^{iw\tau} { \mbox{sgn}(\tau) \over |\tau|^{2\Delta}}= i\, 2^{1-2\Delta}\sqrt{\pi}{\Gamma(1-\Delta)\over \Gamma({1\over 2}+\Delta )}|w|^{2\Delta-1}\mbox{sgn}(w)
\label{MS}
\end{equation}

\subsection{Single k and collective equation}

Given the interaction Lagrangian \eqref{Lintchimomenta}, the Schwinger-Dyson (SD) equations are given by 
\begin{eqnarray}
\label{SDEqn1init}
\Sigma^{aa^\prime}_{k_1k_1^\prime}(t) &=& {J^2A^8\over L^3}  \delta^{aa^\prime}F(k_1)F(k_1^\prime) \sum_{k_2,k_3,k_4,k_2^\prime,k_3^\prime,k_4^\prime}  \delta_{\sum_{i=1}^4 (k_i + k'_i)=0}\,
\prod_{i=2}^4 [F(k_i)F(k'_i)G_{k_ik'_i}\ \ \ \ \\ 
\label{SDEqn2init}
( G^{aa'}_{kk'})^{-1} &=& (G^{(0)aa'}_{kk'})^{-1} - \Sigma^{aa'}_{kk'}(\omega)\,.
\end{eqnarray}
Here $G^{(0)aa'}_{kk'}$ is the free two-point function. In going from the interaction lagrangian to the SD equation we carried out the disorder average. We also assumed that the filter, in momentum space, cuts off the interaction before reaching momentum $\sim {\cal O}(L)$, such that the other chirality (in our conventions) does not participate in the interaction (i.e., we are in the chiral model). Practically, this enforces strict momentum conservation, rather than up to multiples of L. 

We will assume that after disorder averaging both the $\mathrm{SO}(N)$ symmetry and the $\mathbb{Z}_L$ lattice translations are preserved\footnote{Verifying this requires an analysis of stability, which we will not do here.}. We implement this by defining
\begin{equation}
G^{ab}_{k',k}(w)=\delta^{ab}\delta_{k+k'=0} G_k(w).
\label{diagG}
\end{equation}
Under this assumption the SD equation \eqref{SDEqn1init} forces the self-energy $\Sigma(\tau)$ to be diagonal too. Thus it is natural to define
\begin{equation}
 \Sigma^{ab}_{k_1k_1^\prime}(\tau)=\delta^{ab}\Sigma_{k_1}(\tau)\delta_{k_1+k_1^\prime=0}\,.
\end{equation} 
The SD equations \eqref{SDEqn1init}-\eqref{SDEqn2init} become 
\begin{eqnarray}\label{SDEqn1fin}
\frac{1}{G_k(\omega)} &=& {1\over G_{k}^{(0)}(\omega) }-\Sigma_k(\omega)\,, \\
\label{SDEqn2fin}
\Sigma_k(\tau) &=& {A^8J^2\over L^3} \,F(k)^2\prod_{h=1}^{3} \biggl( \sum_{k_h} F(k_h)^2 \,G_{k_h}(\tau)\biggr)\,,
\end{eqnarray}
where we also assumed $F(k)$ is an even function of k. Let us introduce new quantities
\begin{equation}\label{SDdef}
{\tilde G}_k(\tau) \equiv F(k)^2 G_k(\tau)\,,\, \, \, {\tilde G}(\tau) \equiv \sum_k  {\tilde G}_k(\tau)  \,,\,\,\, 
{\tilde \Sigma}(\tau) \equiv F(k)^{-2} \Sigma_k(\tau)\,,
\end{equation}
where we already dropped the momentum subscript in ${\tilde{\Sigma}}$ since this is independent of $k$. The SD equations simplify to
\begin{eqnarray}
\tilde{\Sigma}(\tau) & = & {A^8J^{2} \over L^3} \, \tilde{G}(\tau)^{3} \label{SDEqn1}\\
\tilde{G}(\omega) & = & \sum_{k=-{L/2}+1}^{\frac{L}{2}}\frac{1}{(F^2(k)G_{k}^{(0)})^{-1}(w)-\tilde{\Sigma}(\omega)} \label{SDEqn2}\\
{1\over G_k(w) } &=& {1\over G_{k}^{(0)}(\omega)} - F(k)^2 {\tilde\Sigma}(\omega) \label{2ptG}
\end{eqnarray}
In Appendix \ref{replica} we re-derive this set of equations using the replica method.

We will refer to ${\tilde G}$ and $\tilde \Sigma$ as the collective quantities and to \eqref{SDEqn1} and \eqref{SDEqn2} as the collective equations. Solving them requires the evaluation of the sum \eqref{SDEqn2} to write ${\tilde G}(\omega)$ as a function of ${\tilde \Sigma}(\omega)$. \eqref{SDEqn1} and \eqref{SDEqn2} are then two equations for two functions, one in time and one in frequency, which we can hope to solve. For some choices of $F(k)$ this can be done analytically (in the scaling regime at least), and in other cases we can try and evaluate them numerically. In any case, their basic complexity is not much worse than the original ones in the SYK model. Once we know their solution, we can plug $\tilde \Sigma(\omega)$ into the last equation to compute the momentum 2 point propagator.

It may be instructive to revisit part of our holographic motivation to consider these models at this stage. In terms of a possible bulk interpretation, start by examining the propagator at large $k$. In our case, this is approximately the free propagator with a small correction from $\tilde\Sigma(\omega)$ because $F(k)$ cuts off the coupling with $\tilde\Sigma(\omega)$ at high momenta. In the bulk interpretation, these should correspond to the UV modes near the boundary of $AdS$, interacting with some object living in the bulk interior. As the interaction increases, the momentum modes feel more and more this IR object. In this interpretation, the collective quantities encode the dynamics of whatever macroscopic object we have in the bulk, comprised of the strongly coupled dynamics of the low energy (below the filter scale) modes. They are SYK in nature, although we will see that different filters can give us different scaling theories. The presence of this IR object in the bulk feeds into the high momentum modes like a semi-classical object, correcting their propagators.

\subsection{Solving the collective equations in the continuum}
\label{Rcontinuum}

Since evaluating the sum \eqref{SDEqn2} is complicated, and we are interested in the continuum theory any way, we start discussing the latter here. We think of our model as defined on a circle of fixed size $R$ (which we will think of as large), so that the lattice spacing is $R/L$ and take $L\to\infty$ at the end of the computation. The coordinate around the circle $x=\frac{i}{L}\,R$ will be kept fixed. We will require our filter to cut-off momenta at scales much larger than $1/R$ and much smaller than $L/R$, and kept fixed in the limit $L\to\infty$. In this scaling, many momentum modes participate in the collective dynamics, but we still have effectively a continuum theory for the momentum modes above the filter and below $1/L$. 

The $L\to \infty$ limit allows us to approximate \eqref{SDEqn2} by the integral
\begin{equation}
\begin{aligned}
{\tilde G}(\omega)&=\biggl(\int_{-\infty}^{0} dk +\int_{0}^{\infty} dk\biggr) {1\over \bigl( F^2(k)G_{k}^{(0)}(\omega)\bigr)^{-1}-{\tilde\Sigma}(\omega) }\\
&=\int_{-\infty}^{0} {dk\over \bigl( F^{-2}(k)(-i\omega-E_k\bigr)-{\tilde\Sigma}(\omega) } + 
\int_{0}^{\infty} {dk\over \bigl( F^{-2}(k)(-i\omega+E_k\bigr)-{\tilde\Sigma}(\omega) }\,.
\end{aligned}
\end{equation}
Since ${\tilde G}(\omega)$ and ${\tilde\Sigma}(\omega)$ are both purely imaginary, we obtain
\begin{equation}\label{tildeGexp}
{\tilde G}(\omega)=2i\, {\it{\text{Im}}} \int_{0}^{\infty} {dk \over   F(k)^{-2}(-i\omega+E_k)-{\tilde\Sigma}(\omega) }\,.
\end{equation}
We have assumed that the function $F^{-2}(k)$ increases rapidly enough for large $k$ such that the integral converges and we can replace the cut-off $L$ by infinity. 

The discussion above about the scale of the filter is a little subtle since will be interested in scaling filters of the form 
$F(k)\sim |k|^{-\gamma}$ for a range of $\gamma$. These filters need to be cut off at small $k$ and/or at large $k$, depending on $\gamma$. In position space the filter goes like $(x-y)^{\gamma-1}$. We will approach the issue of the cut-off by examining the integral after the fact.
\begin{itemize}
	\item If the integral converges at large $k$ then we don't need to introduce an additional cut-off in equation \eqref{tildeGexp}, or more precisely, introducing such a cut-off $\Lambda$ will change the results by some negative power of $\Lambda$. However, we may still keep this cut-off, as in \eqref{2ptG}, if we want to.
	\item The integral \eqref{tildeGexp} itself has no problem in low momenta, for finite ${\tilde \Sigma}$, since $F^{-2}(0)=0$. This means that the interaction will effectively mix the low momenta degrees of freedom and set them at some scale defined by the filter. Phrased in another way, one might worry that the process of going from the sum to the integral is not correct. If we were to do the exact sum, then momentum modes around $k\sim 1$ (physical momenta of order $1/R$) would give contributions that scale like $1/{\tilde \Sigma}$ in the limit of large interaction strength. We will see that for the filters that we present below, this is much smaller than the total integral contribution and hence we don't need to worry about anomalous contributions of some global modes.
\end{itemize}

For the scaling low pass filter defined by $F^{-1}(k) = |k|^\gamma$, the integral \eqref{tildeGexp} becomes 
\begin{equation}
{\tilde G}(\omega)=2i\, {\it{\text{Im}}} \int_0^\infty {dk\over k^{2\gamma}(-i\omega+ 2\pi\,k/R)-{\tilde\Sigma}(\omega)} 
\end{equation}
where we already linearised the spectrum as in \eqref{freeR}. Working at fixed low frequency and large $\tilde\Sigma(\omega)$, we can eventually neglect $\omega$ with respect to $k$. We will assume this for now and check for self consistency at the end of the computation. 

Using the variable $k  = \left[-R\,\tilde{\Sigma}(\omega)/(2\pi) \right]^{1 \over 2 \gamma +1} \tilde k$, the above integral becomes
\begin{equation}
  {\tilde G}(\omega)=2i\, {\text{Im}}\left[c_\gamma   \left(\frac{R}{2\pi}\right)^{{1 \over 2 \gamma + 1}}  (-\tilde \Sigma(\omega))^{-{2 \gamma \over 2 \gamma +1}}\right]\,, 
\label{GSigma}
\end{equation}
where $c_\gamma  = \int_{0}^{\infty} {d \tilde k \over k^{2\gamma+1}+1} = {\pi \over 2 \gamma +1} \csc({\pi \over 2 \gamma + 1})$ \footnote{The actual integral we get has the contour from $0$ to $\infty$ at an angle $\theta =  -{\pi i \over 2(2 \gamma +1)}$ in the complex plane. Since there are no poles and the contour at $\infty$ vanishes for $\gamma  >0 $,  we can rotate it back to real axis. One can then use the Formulae $\int_0^\infty {dx \over  1 + x^\nu} = {\pi \over  \nu} \csc({\pi \over \nu}) \mbox{ for } \mbox{Re}(\nu) > 1$.}. The resulting collective equations are solved by
\begin{eqnarray}
  \tilde{\Sigma}(\omega) &= &i\,\mbox{sgn}(\omega)\,\frac{2\pi \kappa^3}{R\,Z}\,{\cal K}(\gamma)\,|\omega|^{6\Delta -1} \quad \text{with} \quad {\cal K}(\gamma)\equiv 2^{1 - 6 \Delta} \sqrt{\pi} {\Gamma(1 - 3 \Delta) \over \Gamma({1 \over 2}+3 \Delta )}\,, \label{sigma-sol} \\
  \tilde G(\tau) &=& Z^{\frac{2\gamma}{2\gamma+1}}\,\frac{R}{2\pi}\frac{\kappa}{|\tau|^{2 \Delta }}\, \mbox{sgn}(\tau)
\end{eqnarray}
provided
\begin{equation}
\Delta = \frac{1+4 \gamma }{2 (1+8 \gamma )}\,, 
\label{Delta}
\end{equation}
$\kappa(\gamma)$ satisfies the constraint
\begin{equation}
\kappa^{\frac{1+8\gamma}{1+2\gamma}}\, 2^{1-2\Delta} \sqrt{\pi} {\Gamma(1 - \Delta) \over \Gamma({1 \over 2}+ \Delta )} = 2 c_\gamma \,\sin({\pi \gamma\over 2 \gamma + 1})\,{\cal K}(\gamma)^{-2\gamma/(2\gamma+1)}\,,
\end{equation}
and $Z$ is identified as
\begin{equation}\label{defZ}
  Z = \left(\frac{A^8 J^2\,R^4}{L^3\,(2\pi)^4}\right)^{-\frac{1+2\gamma}{1+8\gamma}}\,.
\end{equation}
For completeness, we also write the solution to the collective equation in frequency space
\begin{equation}
  \tilde{G}(\omega) = i\,\mbox{sgn}(\omega)\,\frac{\kappa\,R}{2\pi} \, Z^{\frac{2\gamma}{2\gamma+1}}\,2^{1-2\Delta} \sqrt{\pi} {\Gamma(1 - \Delta) \over \Gamma({1 \over 2}+ \Delta )}\,|\omega|^{2\Delta -1}\,,
\end{equation}
with $\Delta$ given in \eqref{Delta}.

\subsubsection{IR contribution}

Here, we check the consistency of our approximations and include further comments on the possible modifications that our scaling filters may require. We will be interested in working in the regime $|\omega|\,R\gg 1$ and fixed. There are three reasons to examine the IR more closely.

First, to check that physical momentum modes close to $1/R$ do not change the conclusions above, we require the magnitude $|{\tilde  G}(\omega)|$ from 
\eqref{GSigma}, which captures the contribution from higher momentum modes within the filter, to be larger than the $|{\tilde\Sigma}(\omega)|^{-1}$ contribution from the lower modes. This amounts to working in the range
\begin{equation}
1\ll R\,|{\tilde\Sigma}(\omega)| \quad \Rightarrow \quad \biggl( {A^8\,R^2\,J^2\over L^3}\biggr)\, (R|\omega|)^2 \gg 1
\end{equation}
This condition is compatible with  $|\omega|\,R\gg 1$ fixed for large enough $J$.

Furthermore, our analysis neglected the $-i\omega$ term in $\tilde{G}(\omega)$, while working in some momentum range satisfying $k^{2\gamma+1}\sim {\tilde{\Sigma}}$. Consistency of both conditions is equivalent to
\begin{equation}
  \frac{k}{R\,|\omega|} \gg \frac{k^{2\gamma + 1}}{R\,|\tilde{\Sigma}(\omega)|} \sim 1
\end{equation}
which can also be satisfied in our desired regime. This is again the statement that many momentum modes, much above physical momenta $1/R$ participate in the collective quantities.

Second, the discussion above is valid at intermediate frequencies, as long as we stay  $\omega \gg 1/R$, at which point we see that momentum is quantized. It could still be that there is an almost continuum spectrum of energies originating from large $N$, but in any case, we expect that this limit to be governed by a different limit of SD equations.

Third, a filter of the form $F\sim 1/|k|^\gamma$ is non trivial in position space. Its Fourier transform is ${\tilde F}=x^{\gamma-1}$. We can work with $\gamma<1$ to obtain reasonable behavior in position space, or we can modify the solution to decay further at some long distance, for example by considering
\begin{equation}
F(k)=k^{-\gamma}e^{-D^2/k^2}\, .
\end{equation}
We will then need to choose $D$ to be small enough. Under this assumption, the discussion above remains the same. We could analogously add a hard UV cut-off to improve the UV convergence.

\subsection{Going to the infinite line}\label{sec:NonCompactSoln}
 
At finite radius $R$, we can always rescale $A$ and $J$ keeping $Z$ fixed, so that the model provides finite, L-independent, results. This is the standard RG approach of keeping the IR fixed and running the UV appropriately. In the following, we investigate whether we can achieve the same finiteness in the non-compact limit $R\to \infty$.
 
In the non-compact limit, finite physical momenta are labelled by $p=\frac{2\pi k}{R}$. Despite the rescaling of momenta, the diagonal contribution to the 2 point function in \eqref{diagG} remains finite
\begin{equation}
G^{\infty}(p) = G_{k=pR/2\pi}\,.
\end{equation}

To investigate the existence of a finite interacting theory in the deep IR in the non-compact limit, we will require both the filter function and the two point function \eqref{2ptG} to be finite. 

The first condition requires to write the filtering function $A\,F(k)= \frac{A}{|k|^\gamma}$ relating the interacting fermions $\eta$ with the physical fermions $\chi$ in terms of the physical momentum $p$ as $A_\infty\,F(p) \equiv \frac{A_\infty}{|p|^\gamma}$. Hence, we learn $A\sim A_\infty\,R^\gamma$, with $A_\infty$ fixed.

The second condition is studied by replacing \eqref{sigma-sol} into \eqref{2ptG}
\begin{equation}
G^{\infty}(p) = {1 \over [G^{(0)}_{k}(\omega)]^{-1} - i\, \mbox{sgn}(\omega) |\omega|^{6 \Delta -1}\,\kappa^3\,\frac{(2\pi)^{2\gamma+1}{\cal K}(\gamma)}{R^{2\gamma+1}\,Z\,p^{2\gamma}}}
\label{finiteGR}
\end{equation}
Requiring the interaction to be finite is equivalent to keeping
\begin{equation}
\frac{ J^2}{\Lambda_0^3} \ \ \mbox{fixed}\,.
\label{nonc}
\end{equation}
where recall that $\Lambda_0 = {L \over R}$ is the inverse lattice spacing. 

\subsubsection{Hyperscaling}

The finite 2 point function \eqref{finiteGR} has an hyperscaling symmetry in the deep IR $\omega\to \lambda^{-z}\omega$ and $p\to \lambda^{-1}\,p$\footnote{The $z$ exponent is usually defined in real space by the scaling relations $t\to \lambda^z\,t$ and $x\to \lambda\,x$.}
with dynamical scaling exponent $z$:
\begin{equation}
  z = \frac{2\gamma+1}{6\Delta - 1} = {1 \over 2} + 4 \gamma\,.
\end{equation}
This covers the range $z>1/2$. In particular, it includes unitary theories, i.e. those with $z\geq 1$. 

The range of $z$'s slightly changes when we consider random couplings of $q$ fermions ($q=4$ in our previous analysis). Then $\Delta_q \equiv {1 + 4 \gamma \over 2(1+2 q \gamma)} $ and the dynamical exponent becomes 
\begin{equation}
  z_q = \frac{2\gamma+1}{6\Delta_q - 1} = {1 + 2 q \gamma \over q-2 }  \,.
\end{equation}
This opens further possibilities for the range of $z$.

\subsection{Continuum non-compact limit}
\label{Raction}

In this subsection we discuss again the continuum model formulated on an infinite line, but from the action perspective. We will recover the condition \eqref{nonc} and in the process we will give the continuum version of the impurity average \eqref{impurity}.

First, let us write the continuum $L\to \infty$ limit of our interacting theory on a circle of size $R$. Using \eqref{freeR}) and \eqref{Lintchimomenta},
\begin{equation*}
\mathcal{L} = \sum_p  \, \chi^{a}_{-p}(\partial_{\tau }+ p )\chi^{a}_p + {(2\pi)^{4\gamma} A^4   \over L^2 R^{4 \gamma} } \sum_{a_i, p_i} \prod_{i=1}^4 [F(p_i) \chi^a_{p_i}] \int {Ldx \over R} e^{- i x \sum_{j=1}^4 p_j}
J_{a_1a_2a_3a_4}(x)\,.
\end{equation*}
Here the sum over momentum $p$ runs from the  IR cutoff $\Lambda_{\text{IR}} \sim {1 \over R}$ to the UV 
cutoff  $\Lambda $. We also defined $F(p) \equiv {1 \over |p|^\gamma}$ as in our discussion in subsection \ref{sec:NonCompactSoln}.

Rewriting the original filter parameter $A$ in terms of the fixed $A_{\infty} =  A ({2 \pi \over R})^{\gamma}$, as in subsection \ref{sec:NonCompactSoln}, the interaction lagrangian becomes 
\begin{equation}
{\mathcal L}_{\text{int}} = { A_\infty^4    \over L R   } \sum_{a_i, p_i} \prod_{i=1}^4 [F(p_i) \chi^{a_i}_{p_i}] \int dx e^{- i x \sum_{j=1}^4 p_j}\,
J_{a_1a_2a_3a_4}(x)\,.
\end{equation}
Lastly, we take the non-compact limit $R \to \infty$. Sums over momentum $\sum_p$ are proportional to $R \int dp$. To keep a finite kinetic term, we need to rescale the physical fermions as $\chi^a_p \sim {\chi^a(p) \over \sqrt R}$, leading to a non-compact lagrangian 
\begin{equation*}
\int dp  \chi^{a}(-p)(\partial_{\tau }+ p )\chi^{a}(p) +{ A_\infty^4 R   \over L   } \sum_{a_i} \int \prod_{i=1}^4 [dp_i F(p_i) \chi^a(p_i)] \int dx e^{- i x \sum_{j=1}^4 p_j}\,J_{a_1a_2a_3a_4}(x)\,.
\end{equation*}

Since the non-compact version of the impurity average \eqref{impurity}
\begin{equation}\label{continuumdisorder}
  E[J_{a_1a_2a_3a_4}(x)\,J_{b_1b_2b_3b_4}(y)] = \frac{3! J^2}{N^3}  \frac{R}{L} \delta (x-y)\,,
\end{equation}
includes an additional $R \over L$ factor from the continuum limit of the discrete Kronecker delta $\delta_{ij}$, we can write the continuum version of the SD equation as
\begin{eqnarray} \nonumber
\Sigma^{a_1a^\prime_1}(p_1,p^\prime_1,\tau) &=&\delta^{a_1a^\prime_1}{ J^2 A_\infty^8 R^3\over L^3} F(p_1) F(p^\prime_1)  \sum_{a_i} \prod_{i=2}^4[\int dp_idp^\prime_i F(p_i) F(p^\prime_i) G^{a_ia_i}(p_i,p^\prime_i,\tau)  ] \cdot \\
&& \hspace{10mm}\cdot \int dx\, e^{-i x \sum_i(p_i + p'_i)}\,, 
\end{eqnarray}
where the last term will implement conservation of momentum $\delta(\sum_i(p_i + p^\prime_i))$. It is now clear that to keep a non-trivial interaction in the non-compact limit we must work with
\begin{equation}
{J^2 R^3 \over L^3} = {J^2 \over \Lambda_0^3} \ \ \ \mbox{  fixed}
\end{equation}
Hence we reproduce our previous claim \eqref{nonc} provided we take the disorder average in the continuum non-compact limit to be as in eq(\ref{continuumdisorder}).

\section{A probe model}

In the previous class of models, we were interpreting the high momentum modes of the physical fermions $\chi^a$ as living outside of a black hole in some putative bulk, while the strongly interacting low momentum modes of $\chi^a$, i.e. the $\eta^a$ degrees of freedom, built the putative black hole. In this section, we explore a second class of models with a similar holographic motivation.

Consider models consisting of two types of fermions : $\eta^a,\ a=1..N$ interacting via an SYK model or its 1+1 extension described in previous sections and a single degree of freedom (or maybe a few) $\rho$ acting as a probe. We envision a situation in which the $\eta^a$ fermions describe the degrees of freedom of a black hole (in some approximate sense), while the $\rho$'s encode the analogue of single trace operators in the AdS/CFT correspondence

More specifically, we will take $\rho$ to be a 1+1 system (but we can take them in any dimension), so that they become $\rho^i$ where $i$ is the spatial index. It can either be fields that go to a free fermion in the continuum, or we can maybe take them to be some generalized free fields, in which case we can hope to find a field of arbitrary dimension.

%

\subsection{A 2D probe model with SYK kernel}

In this subsection we introduce, and solve in some regime, one such model. Let $\eta^{a}$, $a=1,..N$ be the $0+1$d SYK Majorana fermions and $\rho^{i}$ be Majorana fermions on a periodic lattice of length $L$ ($i=0,..L-1$ is the spatial index and $\rho^0\equiv \rho^L$). Following previous sections, we will find it useful to have a filter for the $\rho$ fermions.

We take our model to have the action $S = S_{\eta} + S_{\rho} + S_{\rho,\eta}$ where 
\begin{eqnarray}
S_{\eta} &=&\int d\tau [  {1 \over 2} \eta^a \partial_\tau \eta^a + \sum_{a,b,c,d} J_{abcd}\eta^{a}\eta^{b}\eta^{c} \eta^{d} ]\\
S_{\rho} &=& \sum_i  \int d\tau \{ {1 \over 2} \rho^i \partial_\tau \rho^i - i \alpha [\rho^i , \rho^{i+1} ]  \} = \sum_k \int d\tau \{ {1 \over 2} \rho_{-k} (\partial_\tau +E_k)  \rho_k   \} \\
S_{\rho ,\eta} &=&   \int d\tau \sum_{i_{1}..i_{k}}\hat{J}_{i_{1}..i_{k}a_1..a_m}\bar \rho^{i_{1}}\bar \rho^{i_{2}}..\bar \rho^{i_{k}}\eta^{a_{1}}..\eta^{a_{m}} \\
&=& {A^k \over L^{k \over 2}} \int d\tau  \sum_{i_1,..i_k, k_1,..k_k,a_1..a_m} [\prod_{j=1}^k F(k_j) \rho_{k_j} ]
e^{-{2\pi i \over L} \sum_{j=1}^k i_j k_j  } \ \eta^{a_{1}}..\eta^{a_{m}}  \hat J_{i_{1}..i_{k}a_1..a_m}
\end{eqnarray}
$\rho_k$ stands for the Fourier transform of the $\rho^i$ lattice fermions, whereas $\bar \rho_i$ fermions are the corresponding low pass fermions $\bar \rho_i \equiv {1 \over \sqrt L} \sum_k F(k) \rho_k e^{-2 \pi i k \over L } $ interacting with the SYK fermions $\eta^a$. The $\hat{J}_{i_{1}i_{k}a_1..a_m}$ are taken to be random variables with impurity average
\[
E[ \hat{J}_{i_{1}..i_{k}a_1..a_m}\hat{J}_{i_{1}..i_{k}a_1..a_m}] =\frac{k! m!\hat{J}^{2}}{N^{m}} 
\]
and $E[ J_{abcd} J_{abcd}] = {m! J^2 \over N^3}$ as for SYK fermions.

\begin{figure}
\begin{subfigure}{.5\textwidth} 
\vspace{-50pt}
\includegraphics[scale=0.5]{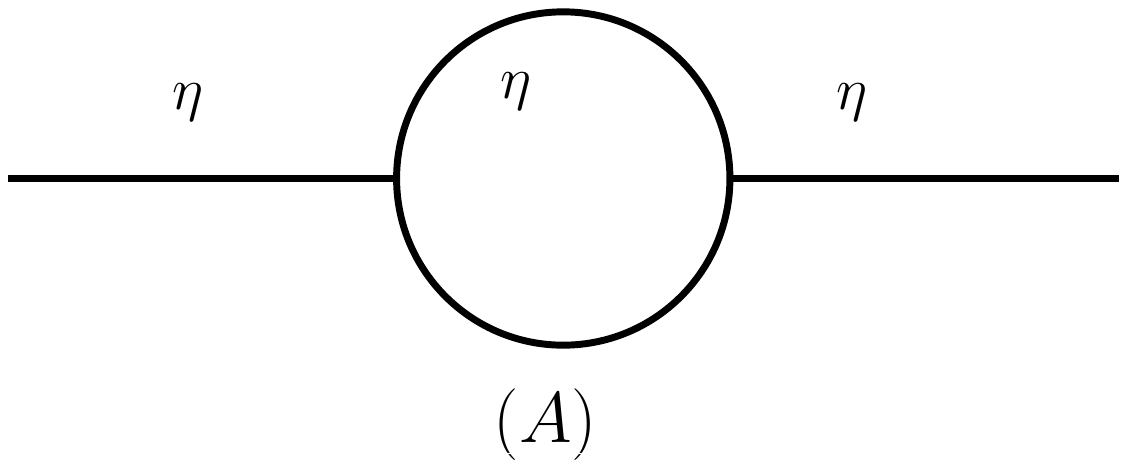}\\
\vspace{-300pt}
\end{subfigure}
\begin{subfigure}{.5\textwidth} 
\vspace{-50pt}
\includegraphics[scale=0.45]{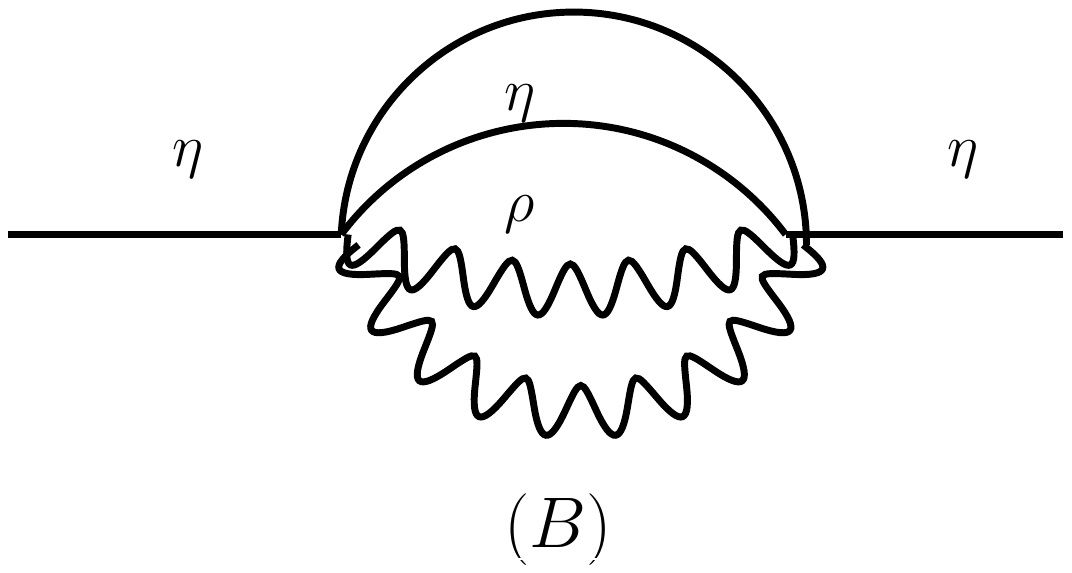}\\
\vspace{-250pt}
\end{subfigure}
\caption{Diagrams contributing to the $\eta$ propagator. }
\label{diag1}
\end{figure}

The $N$ scaling was chosen so that the $\rho^i$ fermions behave like probes, i.e. their propagator will be corrected by the interactions whereas the $\eta^a$ propagators will remain unmodified at leading order. More precisely, there are two leading 1-loop diagrams contributing to the 1PI self-energy
of the $\eta^a$ propagator, as indicated in figure \ref{diag1}. Diagram (A) scales like $E[J^{2}_{..}]N^{3}\sim{\cal O}(N^{0})$ as in SYK. Diagram (B) is subleading since it scales like $E[\hat{J}^{2}_{..}]N^{m-1}\sim{\cal O}(N^{-1})$. Hence, the $\eta^a$ propagators which we denote by $G(\tau)$ are indeed unmodified at leading order.

Next we will solve for the $\rho$ propagator which we will denote by $\cal G$. The leading 1-loop diagram (denoted by $\cal S$) contributing to its 1PI self-energy is given in figure \ref{diag2}.     
\begin{figure}\label{diag2}
\centering
\includegraphics[scale=0.45]{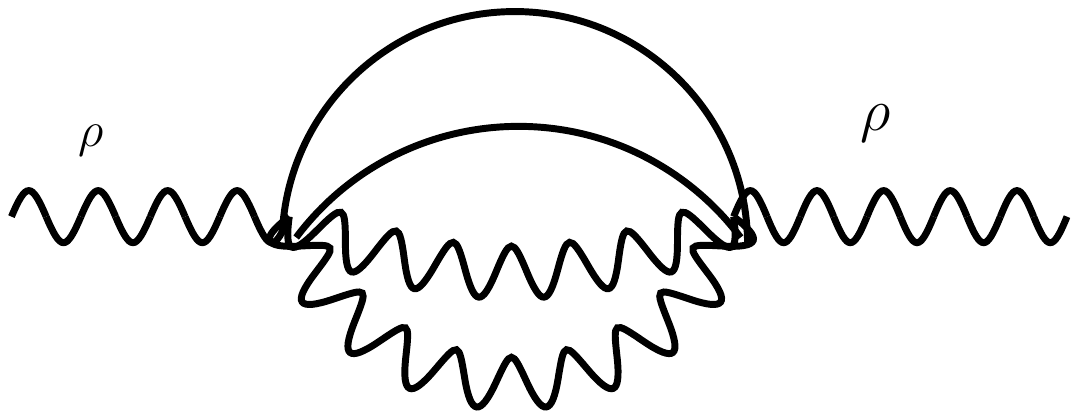}
\vspace{-250pt}
\caption{Diagram contributing to the $\rho$ propagator. }
\label{diag2}
\end{figure}
Since this diagram scales like $\hat{J}^{2}_{..}N^{m}\sim{\cal O}(N^{0})$, it gives rise to a non-trivial correction. In fact we find the SD equations for $\rho$ to be 
\begin{eqnarray} \nonumber 
{\cal S}_{k_1,k_1'}(\tau) &=& {A^{2k} \hat J^2 \over L^k}   G(\tau) ^m   F(k_1) F(k_1')   \times \\
&& \sum_{i_1,..i_k,k_2..k_k,k_2'..k_k'}  \prod_{j=2}^{k} [F(k_j)F(k_j'){\cal G}_{k_jk_j'}(\tau)]  e^{-{2\pi i \over L} \sum_{j=1}^k i_j (k_j+k'_j)   }  \\
{\cal G}_{k,k'}^{-1}(\omega) &=& {{\cal G}^{(0)}}_{k,k'}^{-1}(\omega) - {\cal S}_{k,k'}(\omega)
\end{eqnarray}
It is clear that the SD equations force self energy ${\cal}S$ and hence the propagator ${\cal G}$ to be diagonal in momentum space. We will assume that $F(k)$ is an even function. Let us define
\begin{equation}
{\cal S}_{k,k'}(\tau) \equiv \delta_{k+k'=0}  {\cal S}_k(\tau) \hspace{10mm} {\cal G}_{k,k'}(\tau) \equiv \delta_{k+k'=0} {\cal G}_k(\tau) 
\end{equation}
The SD equations then become 
\begin{eqnarray}
{\cal G}_{k}^{-1}(\omega) &=& {{\cal G}^{(0)}}_{k}^{-1}(\omega) - {\cal S}_k(\omega)   \\
 {\cal S}_k(\tau) &=& A^{2k} \hat J^2 F(k)^2  G(\tau)^m [\sum_{k'} F(k') {\cal G}_{k'}(\tau)]^{k-1}
\end{eqnarray}
We finally write a collective SD equation by defining 
\begin{equation}
{\cal S}_{k}(\tau) \equiv   F(k)^2 \tilde {\cal S}(\tau) \hspace{10mm} \tilde {\cal G}(\tau) \equiv  \sum_k F(k)^{2} {\cal G}_k(\tau) 
\end{equation}
We have dropped the subscript $k$ on $\tilde {\cal S}$ because the SD equations force it to be independent of $k$. The final SD equations become
\begin{eqnarray}
\tilde {\cal G}(\omega) &=& \sum_k {1 \over F(k)^{-2}  {{\cal G}^{(0)}}_{k}^{-1}(\omega) - \tilde {\cal S}(\omega) }\\
\tilde  {\cal S}(\tau) &=& A^{2k} \hat J^2  G(\tau)^m \tilde {\cal G}(\tau)^{k-1} \\
 {\cal G}_{k}(\omega) &=& {1 \over {{\cal G}^{(0)}}_{k}^{-1}(\omega) - F^2(k) \tilde {\cal S}(\omega) }  
\end{eqnarray}

These equations can be solved using the available SYK solutions in the conformal window $G(\tau) \sim {1 \over \tau^{1 \over 2}}$ (see \cite{Maldacena:2016hyu}), and the same strategy we followed in section 3. Rather than presenting the solution for arbitrary $k$ and $m$, we focus on the $k=m=2$ case. 

The scaling of the collective probe propagator and self-energy is
\begin{eqnarray}
{\tilde{\cal G}}(\tau)\sim |\tau|^{-\Delta_1},\ \ {\tilde{\cal G}}\sim |\omega|^{\Delta_1-1}\\
{\tilde{\cal S}}(\tau)\sim |\tau|^{-\Delta_2},\ \ {\tilde{\cal S}}\sim |\omega|^{\Delta_2-1}\\
\end{eqnarray}
with
\begin{equation}
\Delta_2-1=\frac{1+2\gamma}{1+4\gamma},\ \ \ \Delta_1-1=- \frac{2\gamma}{1+4\gamma}\,.
\end{equation}
Notice these are consistent with the assumption $|\omega|\ll {\tilde{\cal S}}$ holding in the deep scaling IR regime. 

\subsection*{Acknowledgements}
We would like to thank S. Ross for collaboration at early stages of this project. The work of MB is supported by an ISF center of excellence grant (1989/14). PN gratefully acknowledges the support from International Centre for Theoretical Sciences (ICTS), India. The work of MR is supported by a Discovery grant from NSERC. The work of JS is supported by the Science and Technology Facilities Council (STFC) [grant number ST/L000458/1].

\begin{appendix}

\section{Equations via the replica method}
\label{replica}
We now rederive the Schwinger-Dyson equations for our model using
replica methods. In this framework those equations represent the saddle point approximation to the effective action of the model. 

In order to compute $S^{(n)}$, the n'th Renyi entropy, we construct $n$ replicas
of our model, labelled by $\alpha=1,\dots n$ The Euclidean action for the replicated theory is
\begin{eqnarray}
S^{(n)}&=&\int d\tau\sum_{\alpha} \left[ \frac{1}{2}\sum_{i,a}\chi_{\alpha}^{i,a}(\tau)\partial_{\tau}\chi_{\alpha}^{i,a}(\tau)  -  i\alpha\sum_{i,a}[\chi_{\alpha}^{i,a}(\tau),\chi_{\alpha}^{i+1,a}(\tau)] \right. \nonumber \\
&&~~~~~~~~~~~~~~~~~~~~~~~~~~\left. + \sum_{i,abcd}J_{i,abcd} \, \eta_{\alpha}^{i,a}(\tau)\eta_{\alpha}^{i,b}(\tau)\eta_{\alpha}^{i,c}(\tau)\eta_{\alpha}^{i,d}(\tau)\right]
\end{eqnarray}
We now perform the disorder average, recalling that we have independent  disorder variables at each site, we get \begin{eqnarray}\nonumber
S^{(n)}&=&\int d\tau\sum_{\alpha,a,i}\left[\frac{i}{2} \chi_{\alpha}^{i,a}(\tau)\partial_{\tau}\chi_{\alpha}^{i,a}(\tau)-i\alpha [\chi_{\alpha}^{i,a}(\tau),\chi_{\alpha}^{i+1,a}(\tau)]\right]\\
&&~~~~~~~~~~  - \frac{4 J^{2}L^3}{N^{3}}\sum_{\alpha,\beta}\int d\tau\int d\tau' \,\sum_i\left[\sum_{a}\eta_{\alpha}^{i,a}(\tau)\eta_{\beta}^{i,a}(\tau')\right]^{4}
\label{ReplicaStart}
\end{eqnarray}
where we use the convention $E(J_{....},J_{....}) \sim \frac{J^2 L^3}{3! N^3}$ for each  randomly distributed variable.

One can now perform a  Hubbard-Stratonovich  transformation by introducing the real decoupling field $Q^{i}_{\alpha\beta}(\tau,\tau')$, symmetric in replica indices, for each site
\begin{eqnarray}
S^{(n)}&=&\int d\tau\sum_{\alpha,a,i}\left[\frac{1}{2} \chi_{\alpha}^{i,a}(\tau)\partial_{\tau}\chi_{\alpha}^{i,a}(\tau)-i\alpha [\chi_{\alpha}^{i,a}(\tau),\chi_{\alpha}^{i+1,a}(\tau)]\right]\\
\nonumber
 &  & \ \ +\sum_{\alpha,\beta}\int d\tau \int d\tau'\sum_i\left[ \frac{N}{4 L  J^2} Q^i_{\alpha\beta}(\tau,\tau')^{2}- \frac{2 L }{ N}Q^i_{\alpha\beta}(\tau,\tau') \left( \sum_{a}\eta_{\alpha}^{i,a}\eta_{\beta}^{i,a}\right)^{2}\right]
\end{eqnarray}
Finally we introduce another set of decoupling fields $P^i_{\alpha \beta}$, also real and symmetric in replica indices, to obtain
\begin{eqnarray}
S^{(n)}&=&\int d\tau\sum_{\alpha,a,i}\left[\frac{1}{2} \chi_{\alpha}^{i,a}(\tau)\partial_{\tau}\chi_{\alpha}^{i,a}(\tau)-i\alpha [\chi_{\alpha}^{i,a}(\tau),\chi_{\alpha}^{i+1,a}(\tau)]\right] \nonumber\\
\nonumber
 &  & \ \ +\frac{N}{L} \sum_{\alpha,\beta}\int d\tau \int d\tau' \sum_i\left[ \frac{Q^i_{\alpha\beta}(\tau,\tau')^{2}}{4 J^2} +\frac{Q^i_{\alpha\beta}(\tau,\tau') P^i_{\alpha\beta}(\tau,\tau')^2}{2} \right.\\
 && \left.\hspace{40mm}-  Q^i_{\alpha\beta}(\tau,\tau') P^i_{\alpha \beta}(\tau',\tau)  \left(\frac{L}{N} \sum_{a}\eta_{\alpha}^{i,a}\eta_{\beta}^{i,a}  \right) \right]
\end{eqnarray}

This is simply the sum over the replicated action obtained for each site separately, for a direct comparison see for example the discussion in \cite{Sachdev:2015efa,Fu:2016yrv}. Note that the saddle point equations set 
\begin{eqnarray}
P^i_{\alpha \beta}(\tau,\tau')&=&\, \frac{L}{N }\sum_a \langle \eta^i_\alpha (\tau)\eta^i_\beta(\tau') \rangle \nonumber \\
Q^{i}_{\alpha \beta}(\tau,\tau')&=&  {J^2} P^i_{\alpha,\beta}(\tau,\tau')^2
\end{eqnarray}

We now assume that replica symmetry is not broken, so that $P^i_{\alpha\beta}=P^i \delta_{\alpha\beta}$ and   $Q^i_{\alpha\beta}=Q^i
\delta_{\alpha\beta}$. Similarly we assume that upon disorder averaging the $SO(N)$ symmetry is restored. Therefore we can drop the fermion $SO(N)$ index and refer to a single fermion. We further can go to a single replica, obtaining the action
\begin{eqnarray}
S &=&N \int d\tau\sum_{i}\left[\frac{1}{2} \chi^{i}(\tau)
\partial_{\tau}\chi^{i}(\tau)-i\alpha
[\chi^{i}(\tau),\chi^{i+1}(\tau)]\right]
\\
\nonumber
 &  & \ \ +\frac{N}{L} \int d\tau \int d\tau' \, \sum_i\left[
\frac{Q^i(\tau,\tau')^{2}}{4 J^2} +\frac{Q^i(\tau,\tau') P^i(\tau,\tau')^2}{2}
 -  L \, Q^i(\tau,\tau') P^i(\tau',\tau)   \eta^{i}(\tau)\eta^{i}(\tau')
\right]
\end{eqnarray}
We are now ready to integrate out the fermions $\chi^{i}$. Define the mass shifts $\tilde \Sigma^i= Q^i P^i$, making the same self-consistent assumption as above, namely that the mass shifts are on-site only (i.e. they are all equal in momentum space), so that the saddle point solution satisfies $\tilde \Sigma_k=\tilde \Sigma$, i,.e the same value for each Fourier mode $k$. With this assumption we can now Fourier transform the action and integrate out the fermions:
\begin{eqnarray}
S&=& N  \, \sum_{\omega,k} \log \text{Pf} \left[ \partial_{\tau} - E_k - \tilde \Sigma(\tau,\tau') F(k)^2\right]  \\
\nonumber
 &  & ~~~~ +{N} \sum_{k}\int d\tau \int d\tau' \, \left[  \frac{Q_k(\tau,\tau') Q_{-k}(\tau,\tau')}{4 J^2} + \frac{P_{k}(\tau,\tau') \, \tilde \Sigma (\tau,\tau') }{2} \right]
\end{eqnarray}
where we have used that
 in momentum space $\eta_k=F(k) \chi_k$.  We can further use the identity $Q^i= J^2 (P^{i})^2$, and denote $\tilde G^i=\frac{P^i}{L}$ to obtain
 \begin{eqnarray}
S&=& N  \, \sum_k \log \text{Pf} \left[ \partial_\tau - E_k - \tilde \Sigma(\tau,\tau') F(k)^2\right]  \\
\nonumber
 &  & ~~~~~~~~ +{N}\sum_{k}\int d\tau \int d\tau' \, \left[  \frac{J^2 (\tilde G(\tau,\tau')^2)_k (\tilde G(\tau,\tau')^2)_{-k}}{4 L^4 } + \frac{ \tilde G_{k}(\tau,\tau') \, \tilde \Sigma (\tau,\tau')}{L}\right]
\end{eqnarray}
where $(\tilde G(\tau,\tau')^2)_k$ denotes the Fourier transform of $\tilde G^i(\tau,\tau')^2$.

We can now obtain the saddle point equations following from the action. Varying with respect to $\tilde \Sigma$ gives (in frequency space)
\begin{eqnarray}
\tilde G_k(\omega) = {1 \over -i \omega -E_k -\tilde \Sigma(\omega) F(k)^2}
\end{eqnarray}
whereas varying with respect to $G(\tau)=\sum_k\tilde G_k(\tau)$ gives
\begin{equation}
\tilde \Sigma(\tau)= \frac{ J^2}{L^3} \tilde G(\tau)^3 \end{equation}

To exhibit the dependence on the normalization $A$, we redefine $F(k)\rightarrow A F(k)$ and rescale $\tilde{\Sigma}\rightarrow A^{-2} \tilde \Sigma$ and $\tilde G\rightarrow A^{2} \tilde G$. This yields the same equations as those derived in subsections \ref{Rcontinuum} and \ref{Raction}, where the normalization constant $A$ is shown explicitly.

\section{Gaussian low pass filter}
\label{sgaussian}

In this Appendix, we consider the gaussian low pass filter. Although we will not be able to solve the SD equations exactly (even in the deep IR), we will determine the scaling of the 2 point function in frequency space $()\omega)$. Recall the gaussian low pass filter is defined by the function
\begin{equation}
\label{gaussian}
F(k)=e^{-\pi^2 k^2\over R^2 D^2}\,,
\end{equation}
where $D = {\hat D \over R}$ is the physical scale of the filter as can be seen by taking the non-compact limit $R\to \infty$ keeping the physical momentum $p\sim \frac{k}{R}$ fixed.

If we were to consider an step function filter, one would expect to obtain similar physics to the SYK model for the modes passing the filter, while decoupling those being filtered out. What we show below is that the gaussian filter model provides logarithmic corrections to the SYK scaling behaviour for very long times.

To solve this model in the same regime as we discussed the solution for the power law filter, we need to consider the integral \eqref{tildeGexp} with $F(k)$ given by \eqref{gaussian}. This looks like (dropping the $-i\omega$ term compared to $\Sigma(\omega)$  in the deep IR)
\begin{equation}
\tilde G(\omega) = {2 i} \mbox{ Im} \int_0^\infty {d \tilde k  \over e^{\pi^2 \tilde k^2} \tilde k  - {\tilde \Sigma(\omega) \over R D }  } =   2i\,\frac{|{\tilde\Sigma(\omega)| }}{RD}\,\int_0^\infty {d \tilde k   \over \tilde k^2\,e^{2\pi^2 \tilde k^2}  + \frac{|{\tilde\Sigma(\omega)}|^2}{R^2\,D^2} }
\end{equation}
where we used that $\tilde \Sigma(\omega)$ is purely imaginary. Although we could not solve the above integral exactly, we can estimate it for large values of 
$\frac{|{\tilde\Sigma(\omega)|}}{RD}$. In this regime, the integral cuts off when the two terms become comparable. This occurs around $k \sim \sqrt{\log
\frac{|\tilde\Sigma(\omega)|}{RD}}$. Thus we get the following estimate
\begin{equation}\label{gaussianSD2}
|\tilde{G}(\omega)| \sim \frac{\sqrt{\log \frac{|\tilde\Sigma(\omega)|}{RD}}}{\frac{|\tilde\Sigma(\omega)|}{RD}}\,,
\end{equation}
The other SD equation \eqref{SDEqn1} becomes
\begin{equation}\label{gaussianSD1}
{\Sigma(\tau) \over R D}  = {A^8 J^2 \over R D L^3} G(\tau)^3
\end{equation}
Let us now assume a simple ansatz $\tilde \Sigma(\omega) \sim (\omega)^\alpha |\log \omega|^\beta$. Since we will work in small $\omega$ and large time $t$ we can use the following approximation when performing the Fourier transform 
\begin{eqnarray}\nonumber
\int d\omega e^{i \omega t} (\omega)^\alpha |\log \omega|^\beta &\sim&    {(\log t)^\beta \over t^{1 +\alpha }} \left[\int_{-\infty}^\infty d\hat \omega e^{i \hat \omega} \hat \omega^\alpha \left( 1 -  {\beta |\log \hat \omega | \over(\log t)^\beta } + {\cal O}(\log t)^{-2\beta} \right)  \right] \\
&=& {(\log t)^\beta \over t^{1 +\alpha }} \left( 1 + + {\cal O}(\log t)^{-\beta}  \right) 
\end{eqnarray}
The situation is thus very similar to the original SYK model, except for the extra $\log$ pieces. Defining ${\cal J}^2 \equiv  {A^8 J^2 \over R D L^3}$, one can check that the SD equations \eqref{gaussianSD1} and \eqref{gaussianSD2} are solved by
\begin{equation}
|\tilde{G}(\omega)|\sim {\cal J}^{-1\over 2}\,\frac{|\log (\omega/{\cal J}) |^{\frac{1}{8}}}{|\omega/{\cal J}|^{\frac{1}{2}}}\,, \quad\quad |\tilde{\Sigma}(\omega)|\sim {\cal J}^{\frac{1}{2}}|\omega/{\cal J}|^{\frac{1}{2}}|\log (\omega/{\cal J})|^{\frac{3}{8}}\,.
\end{equation}
or in euclidean time $\tau$
\begin{equation}
|\tilde{G}(\tau)| \sim {\cal J}^{-1/2}\,\frac{\log |{\cal J}\tau|^{\frac{1}{8}}}{|{\cal J}\tau|^{1/2}}\,,\quad \quad |\tilde{\Sigma}(\tau)| \sim {\cal J}^{1/2}\,\frac{ \log |{\cal J}\tau|^{\frac{3}{8}}}{|{\cal J}\tau |^{\frac{3}{2}}}\,.
\end{equation}
We see that the resulting theory has $\log (\omega)$ enhancement compared to SYK in the free energy.

\end{appendix}

\bibliographystyle{JHEP}
\bibliography{bibl}

\end{document}